# Ultrafast all-optical switching using doped chromoprotein films

**Szilvia Krekic,[1,2,3] Mark Mero,[4] András Dér,[1]*, and Zsuzsanna Heiner[3]***

[1]Institute of Biophysics, Biological Research Centre, Temesvári krt. 62, 6726, Szeged, Hungary

[2]Doctoral School of Multidisciplinary Medical Sciences, University of Szeged, Dugonics tér 13, 6720, Szeged, Hungary

[3]School of Analytical Sciences Adlershof, Humboldt-Universität zu Berlin, Albert-Einstein-Straße 5-11, 12489 Berlin, Germany

[4]Max Born Institute for Nonlinear Optics and Short Pulse Spectroscopy, Max-Born-Straße 2a, 12489 Berlin, Germany

*Corresponding authors' e-mail addresses: heinerzs@hu-berlin.de, der.andras@brc.hu

## Abstract

Next-generation communication networks require > Tbit/s single-channel data transfer and processing with sub-picosecond switches and routers at network nodes. Materials enabling ultrafast all-optical switching have high potential to solve the speed limitations of current optoelectronic circuits. Chromoproteins have been shown to exhibit a fast light-controlled refractive index change much larger than that induced by the optical Kerr effect due to a purely electronic nonlinearity, alleviating the driving energy requirements for optical switching. Here, we report femtosecond transient grating experiments demonstrating the feasibility of < 200-fs all-optical switching by hydrated thin films of photoactive yellow protein, for the first time, and compare the results with those obtained using bacteriorhodopsin. Possibilities for the practical utilization of the scheme in extremely high-speed optical modulation and switching/routing with nominally infinite extinction contrast are discussed.

## Introduction

The exponential growth of Internet traffic carrying optically coded data trains (packets) requires telecommunication channels with Tbit/s rate and a corresponding sub-picosecond switching speed.[1–5] At Internet nodes, where high switching speed and repetition rate are especially important, currently so-called optical transceiver modules are used as electrical-optical interfaces with accompanying multiple conversions between optical and electric coding along an information channel. All-optical solutions are preferred to avoid any unnecessary delay in signal processing.[3,6] Modulation and routing can also utilize nonlinear optical (NLO) materials exhibiting large, rapid light-induced change in their absorption or refractive index.[7] Besides testing relatively well-known NLO solid-state materials, such as silicon, lithium-niobate[8], chalcogenide glasses,[9] and wide-bandgap semiconductor waveguides[5], engineering of novel 2D materials[10] is also underway, but still far from practical applications. Another promising line of research is based on synthetic dyes of large, delocalized pi-electron systems showing high third-order polarizability typically without two-photon absorption losses, albeit with sensitivity or stability issues.[11,12] In contrast, natural pi-conjugated biomaterials with chromophores embedded in a stabilizing protein core exhibit high light sensitivity, long-term stability, and are readily available.[13–15] After photon absorption by the chromophore, the protein undergoes a cyclic reaction series called photocycle, where consecutive conformational changes occur through various intermediate states with characteristic absorption spectra before returning to the initial state. Since the first steps in the photocycle often proceed on a sub-picosecond time scale, they potentially can be exploited for ultrafast optical switching.

A nontrivial technical challenge in the application of chromoproteins as NLO materials in all-optical switching is the build-up of stable protein-substrate structures with controlled relative humidity that still allow the photocycle to take place. Doped chromoprotein films combined with passive integrated optical structures represent a convenient solution and have been shown to maintain stable optical properties for years.[16] Since the relative humidity in such hybrid structures affects the protein micro-environment, an in-depth characterization of the NLO properties and photocycle kinetics is mandatory.

One of the most thoroughly investigated chromoproteins for all-optical processing applications is the light-driven proton pump, bacteriorhodopsin (BR) - a membrane protein embedded in quasi-crystalline lipid-protein patches. BR has been used in a wide range of photonic and opto-electronic applications.[16–23] When used as the main adlayer component of integrated optical (IO) waveguides, thin films of BR were shown to be appropriate for photo-induced switching down to the sub-picosecond time domain.[24,25] Recently, photoactive yellow protein (PYP) has emerged as a candidate for all-optical switching applications.[26] In contrast to BR, PYP is a small, water-soluble protein offering much more straightforward incorporation into passive structures. Accordingly, the NLO properties of PYP films have recently been characterized[15,26] and shown promise for use in integrated optical modulators.[27] Nevertheless, utilization of the sub-picosecond photoreaction of the PYP photocycle[28,29] for ultrafast all-optical switching has not been accomplished so far.

The photocycles of BR and PYP in solution are depicted in Scheme 1, both showing fast, sub-picosecond to picosecond initial transitions. While the ultrafast kinetics of BR is well studied,[30–33] the ultrafast absorption kinetics of PYP in solution and crystalline form were only recently investigated.[28,34,35] At the same time, the role of environmental variables and structural details of newly engineered chromophore analogues on the ultrafast photocycle of PYP is still a subject of intensive research.[35–37] The existence of sub-picosecond initial transitions in dried, doped PYP films also needs confirmation.

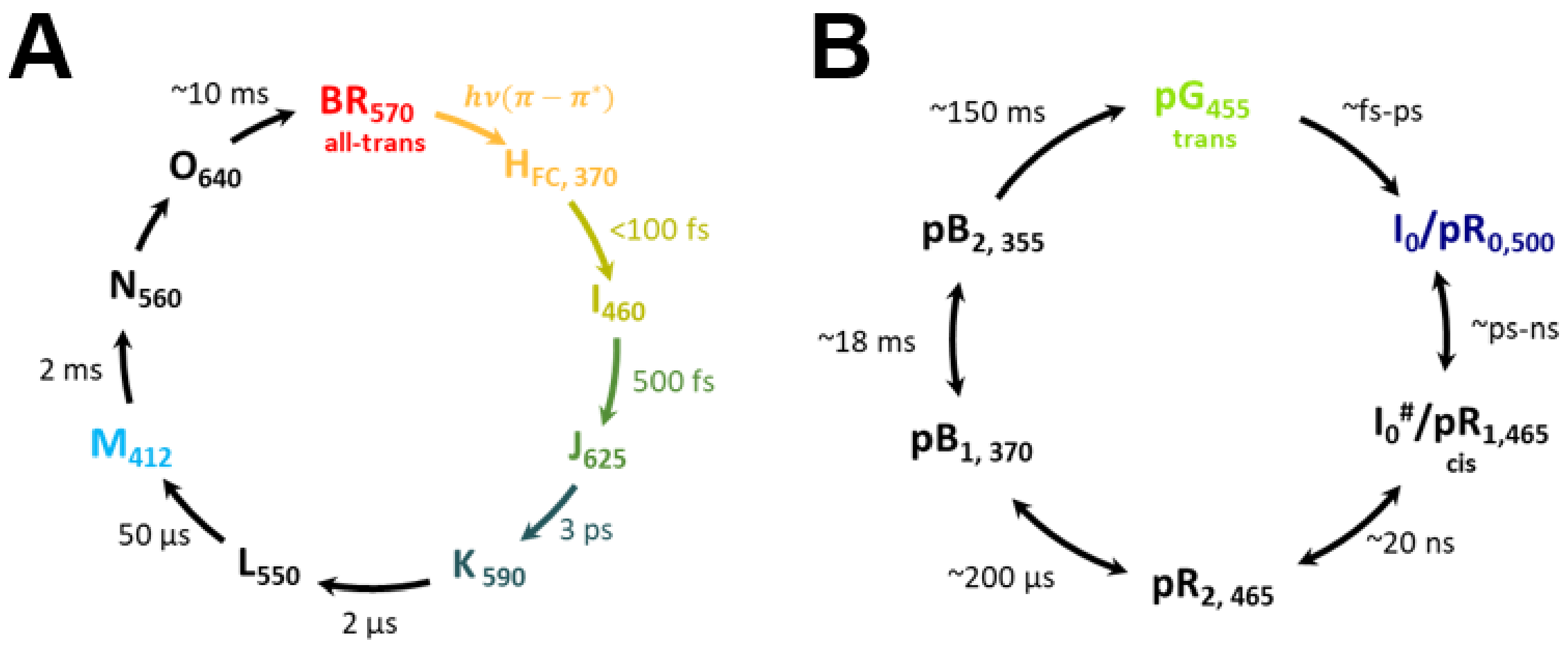

**Scheme 1.** Photocycle model of (A) bacteriorhodopsin and (B) photoactive yellow protein in solution based on Refs. [16,31,34,38–44].

Employing transient grating (TG) spectroscopy,[45,46] we show here that the photocycle of thin, doped PYP films with controlled humidity does contain a sub-picosecond initial step. We follow light-induced conformational changes over the first few ps after excitation, as revealed by the delay-dependent diffracted signal, and compare them with those obtained using thin, doped BR films. Harnessing the TG scheme itself, we successfully exploit the ultrafast kinetics of PYP films for the optical gating of a laser pulse with a rise and fall time down to < 200 fs depending on the film dopant material. The TG scheme naturally offers an arrangement for all-optical switching and modulation, where the diffracted signal represents the optically gated signal with an infinitely high nominal extinction ratio. Our experiments prove the feasibility of ultrafast all-optical switching using the early light-induced molecular transitions of PYP films. Further possibilities towards the practical utilization of such molecular transitions for ultrafast optical switching are briefly discussed.

## Materials and methods

### Sample preparation

For maintaining the water content of the thin protein films, polyacrylamide (PAM) and/or glycerol as doping materials were used. To reduce the cracking of the BR and PYP films, which occurs during the drying phase of the preparation, the protein solutions were first mixed with 87% glycerol as ballast material in various amounts, as explained in Ref[26]. Adding glycerol (GL) enables the creation of thin films in optical quality while also maintains the sample at an appropriate humidity, which is necessary for the integrity of the photocycle of the proteins. Adding glycerol to BR films secures the relative humidity inside the sample at 80-85%, based on our previous optical multichannel analyzer experiments.[26] During sample preparation with GL doping, the initial protein mixtures consisted of 10% volume ratio of glycerol for BR, and 2% for PYP. The solutions were sonicated for 1-2 minutes, then pipetted onto a 160-µm thick BK7 microscope cover slide. The samples were dried under an extractor fume for at least 12 hours before measurements, then they were sandwiched by using a 160-µm thick spacer between two slides to maintain appropriate sample thickness. The optical densities of the GL-doped films were 0.35 and 0.3 for BR and PYP film at our pump wavelength of 515 and 450 nm, respectively. The PAM-doped films were prepared based on a protocol published in Ref[47]. As a result, an 800-µm thick, 7.5 w/w % polyacrylamide layer containing the chromoprotein in high concentration was obtained and subsequently air-dried on a glass surface to form a 220-µm thick, optical-quality film (OD = 0.3 at 450 nm). All TG measurements were conducted at 23°C and at a relative humidity of 30-35%.

## Experimental setup

For our ultrafast TG experiments, the visible optical pump and probe pulses were generated from the output of a commercial Yb laser/amplifier system, which delivered 180-fs laser pulses at a center wavelength of 1028 nm with a repetition rate tunable up to 100 kHz. Pulses at a center wavelength of 514 nm were obtained by second-harmonic generation from the Yb laser. Another portion of the 1028-nm pulses were used for pumping an optical parametric amplifier, which generated pump/probe pulses at wavelengths different from 514 nm. In the experiments discussed below, 450 nm and 480 nm pulses were employed. All pump and probe pulses had a pulse duration of 160 fs. Temporal pulse characterization was performed directly before the TG experiments based on the transient-grating frequency-resolved optical gating (TG-FROG) technique[48,49] by using the same TG setup with a 1-mm thick fused silica plate as the sample.

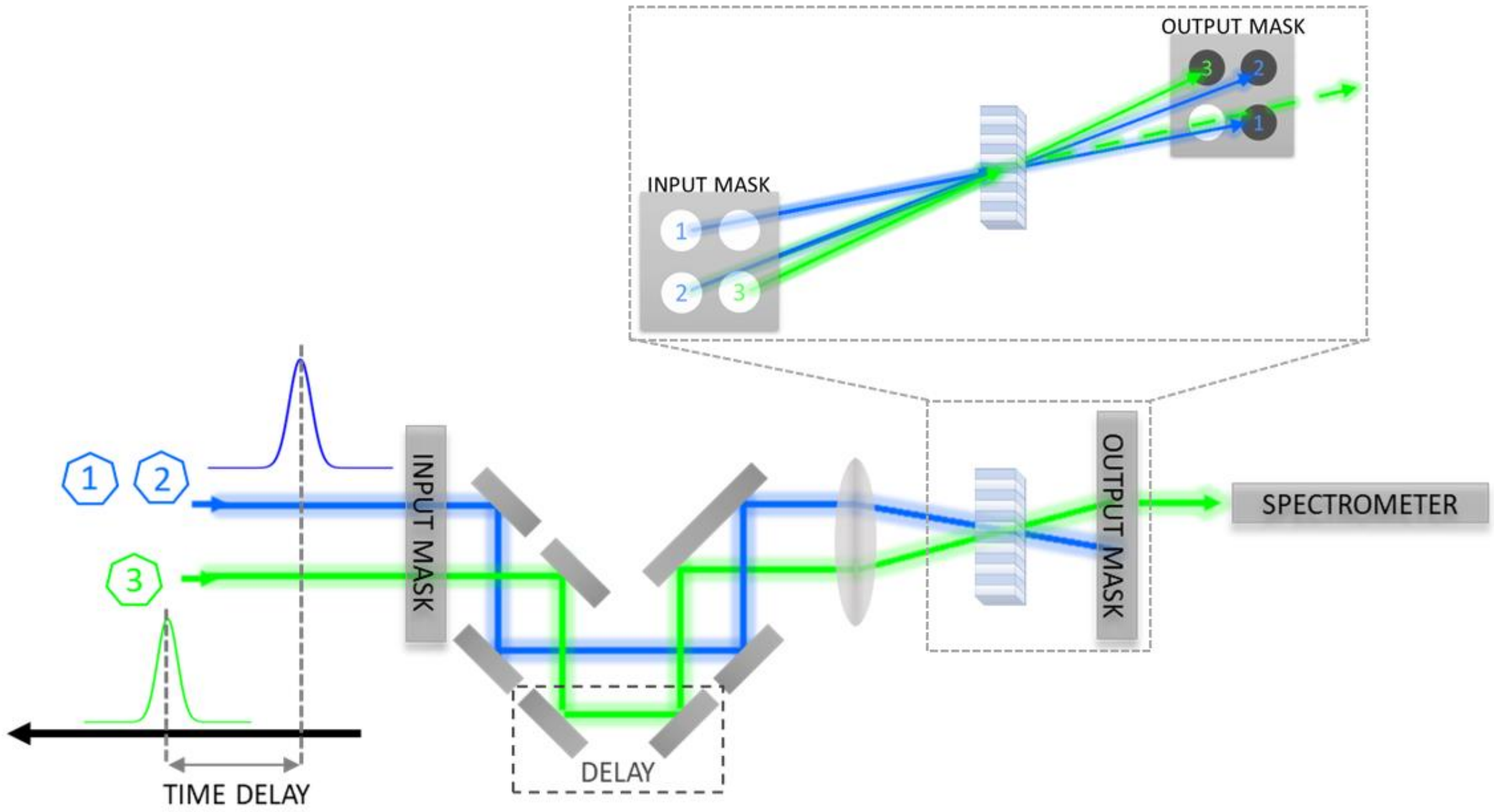

**FIG. 1.** Representative scheme and setup for the ultrafast transient grating experiments.

The schematics of the experimental method and setup are shown in Fig. 1. We used a simplified version[49] of the so-called folded BOXCARS geometry.[50,51] After beam expansion and spatial filtering (not shown), flat-top beams with a diameter of 2 mm were generated by sending up-collimated beams through an input mask. Beamlet 1 and 2 (pump) were derived from one of the expanded input beams and were reflected off the same mirrors securing identical arrival times at the sample. Beamlet 3 (probe) was obtained by transmitting the other input beam through a third hole on the mask and was delayed in time at 20-fs steps relative to Beamlet 1 and 2. An $f$ = 50-mm achromatic lens was used to spatially combine the three pulses (Beamlet 1-3) at the sample, which was placed in the focal plane. Depending on the experiment, the pulse energy per beamlet incident on the sample was 1-12.5 nJ at a repetition rate of 1 Hz. The diffracted signal was collected in the phase matching direction of $\vec{k}_1 - \vec{k}_2 + \vec{k}_3$, where $\vec{k}_{1-3}$ are the wave vectors of Beamlet 1-3, respectively. An output mask was used to spatially separate the diffracted signal from the incident beamlets, which was then detected by a miniature fiber optic spectrometer as a function of delay time. The pump and probe wavelengths were chosen to overlap with the absorption bands of the ground state and the fastest intermediates, respectively. Accordingly, 450-nm and 480-nm pump and 514-nm probe pulses were used for PYP, while 514-nm pump and 450-nm probe pulses were employed for BR (cf. Scheme 1). The two pump wavelengths of 450 and 480 nm in the case of PYP samples were used to test the hypothesis that the kinetics is different at these excitation wavelengths.[52] To induce ground state bleaching of the samples, i.e., creating a steady state that differs from the ground state, additional continuous wave (CW) excitation was used at 405 nm and 532 nm for PYP and BR, respectively.

Data were acquired in the delay range of -0.8 – 4 ps for BR and -0.8 – 10 ps for PYP samples. At each delay, only one single-shot spectrum was recorded and no averaging was performed for any of the time-dependent TG signal curves presented below. Pump-probe scans were repeated at least three times and showed high reproducibility. The background spectrum collected at a delay of -1.8 ps was subtracted from the data. The zero-delay was determined by generating a non-resonant Kerr-type diffraction signal from the uncoated glass substrate of the sample (i.e., no protein present), at an increased pump intensity. Since the spectral bandwidth of the pump and probe pulses was relatively narrow in this study, only a limited range of pixels were used and summed up at each delay to generate the time-dependent diffraction signal for further analysis.

## Results and discussion

### Bacteriorhodopsin

Glycerol-doped thin films of BR were used as our benchmark sample, since its photocycle is well known. The probe wavelength of 450 nm was chosen to fall close to the absorption peak of the I intermediate state of the protein (cf. Scheme 1A). Figure 2A shows the obtained diffracted probe intensity as a function of time (magenta line). The signal reaches its first maximum at the trailing edge of the excitation pulse and decays within 1 ps before it rises again. CW excitation of the sample at 532 nm almost completely diminishes the diffraction signal (cf. Fig. 2A, green line). Signal bleaching is due to the CW excitation creating a steady state population in the rate-limiting, blue-shifted M intermediate state (cf. Scheme 1) with an absorption maximum at 412 nm,[53] which does not have absorption at the applied pump wavelength of 514 nm.[54–56] Therefore, the bleached sample area cannot be excited by the pump laser and a diffraction grating cannot be formed.

To unravel the underlying dynamics, the time-dependent diffracted probe intensity was modelled using the known photocycle scheme for the early intermediates of BR.[25,31,57] Following a 160-fs excitation pulse at 514 nm, only the BR, H, I, J, and K coexisting conformational states of the protein are formed within 4 ps. Accordingly, our model includes the intermediate states and transitions, where [BR], [I], [J], and [K] represent the concentration of the intermediates, and [H] is the concentration of the Franck-Condon state,[31–33] $k_{1\text{-}4}$ are the rate constants, $I_{PUMP}$ is the applied pump intensity, and σ is the absorption cross-section of the ground state.

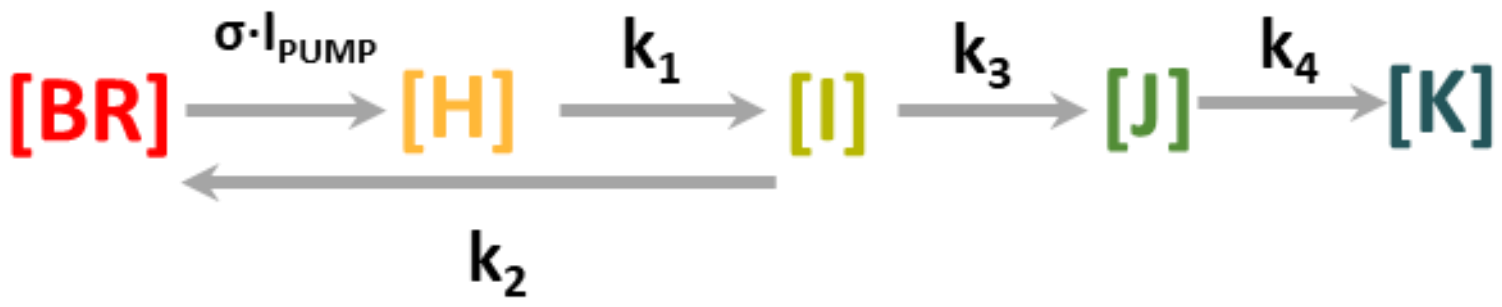

**Scheme 2** Kinetic scheme for rate equations describing the initial steps of the photocycle of bacteriorhodopsin.

The diffracted probe intensity, $\eta(t)$, was assumed to be proportional to the sum of the squares of the peak change in the real and the imaginary parts of the complex refractive index across the induced excited-state concentration grating, which in turn were assumed to be proportional to the concentration of the intermediate state normalized to the initial concentration of the ground state,[46,58]

$$\eta(t) \propto \left[a_i \sum_i \Delta n_i(t)\right]^2 + \left[a_i \sum_i \Delta\kappa_i(t)\right]^2 \propto \left[a_i \sum_i {N_i(t)}/{N_0}\right]^2, \quad (1)$$

where $\Delta n_i(t)$, $\Delta\kappa_i(t)$, and $N_i(t)$ are the change in the real and the imaginary parts of the complex refractive index, and the concentration of state *i*, respectively. Here, state *i* correspond to any state that contributes to the measured signal at the probe wavelength and $a_i$ is a constant. $N_0$ denotes the dark-adapted concentration of the ground state. The time-dependent signal was modelled by calculating the sum of normalized concentration squared values for each time delay through solving the coupled differential equations for the concentrations and employing the least squares method. An absorption cross-section of $\sigma = 0.76\times10^{-16}$ $cm^2$ was assumed[25] and $I_{PUMP}$ was calculated from experimental values. The rate constants were fitted starting from initial values obtained from the literature on BR.[25,31] The contribution of the J intermediate state to the diffracted signal was assumed to be negligible (i.e., $a \approx 0$ in Eq. (1)) due to the fact that the wavelength difference between the 625-nm absorption peak and the 450-nm probe wavelength is much larger than the width of the J-band. Figure 2B shows the so-obtained time-dependent concentrations, while the measured and modelled diffracted probe signals are presented in Fig. 2C.

The above results indicate a clear dominance of the BR-I transition in the measured signal within the first few 100 fs. On the other hand, after a delay of a few ps, the K intermediate state dominates. The rate constants we obtained by modeling the experimental data are $k_1$ = 1/100 $fs^{-1}$, $k_2$ = 1/1500 $fs^{-1}$, $k_3$ = 1/649 $fs^{-1}$, and $k_4$= 1/1787 $fs^{-1}$. These rate constant values are in good agreement with those published earlier.[25,30,31,59–64] Thus, the obtained transient grating signal can be fully reproduced by the standard photocycle model of BR (based on Scheme 2), confirming our hypothesis.

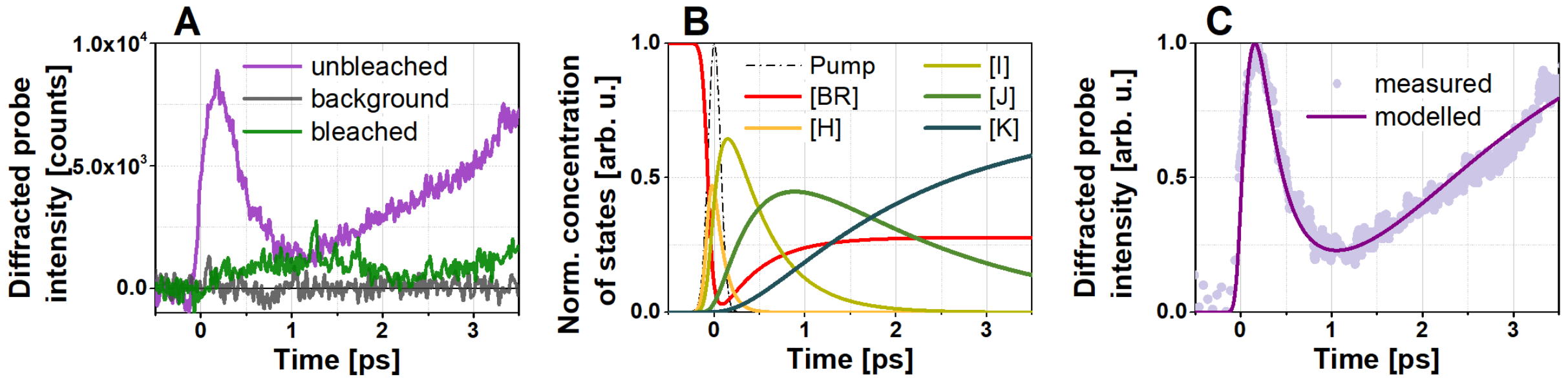

**FIG. 2.** Dynamics of glycerol-doped thin BR films. **(A)** Diffracted probe intensity as a function of time at a probe wavelength of 450 nm. The 160-fs pump pulses were centered at 514 nm. **(B)** Simulation of the normalized concentrations of the H, I, J and K intermediates after 160-fs photoexcitation at 514 nm. The temporal shape of the pump pulse is shown by the dash-dot line. **(C)** Measured and calculated diffracted probe intensity as a function of time.

## Photoactive yellow protein

Measurement of the time-dependent diffraction signal was performed first on glycerol-doped PYP films pumped at 450 nm, which is an excitation wavelength at the peak of the ground state absorption and is most commonly used in the literature. In contrast to BR, the probe wavelength (i.e., 514 nm) was on the red wavelength side of the ground state, pG, to monitor the fastest known intermediate state, which exhibits an absorption peak near 500 nm. The corresponding measured diffraction signal up to a delay of 10 ps is shown in Fig. 3A (cf. orange line). The signal reaches its maximum at a delay of 140 fs and decays to 50% and 10% of its peak value at 520 fs and 1.6 ps, respectively, without any recovery over the measured delay range of 10 ps.

As in the experiments on BR, we attempted quenching of the diffracted transient grating signal by bleaching the ground state using CW excitation at 405 nm. While this quenching does, indeed, decrease the population of the ground state, at the same time, it increases the populations of all other states, including $pR_{0,1,2}$ and $pB_{1,2}$ states. Due to their smaller rate constants (longer decay times), mainly the blue-shifted pB states are expected to form as in the case of the M intermediate in the BR photocycle. Since the wavelength of the exciting laser is not totally outside of the absorption band of the pB states[65] (cf. Scheme 1), it can re-excite PYP molecules in the pB states, driving a portion of them back to the ground state.[66] Accordingly, the diffracted probe intensity decreased in proportion to the bleaching intensity but did not completely vanish, as shown in Fig. 3A (red, purple, blue lines). For example, at an average bleaching power of 30 mW, corresponding to an intensity of 1.0 W/cm$^2$, the diffracted signal decreased by 91 %. Using a different bleaching wavelength, for example, shifted more towards the red edge of the PYP ground state absorption, the diffraction is expected to be fully eliminated. Overall, our results here demonstrate sub-picosecond all-optical switching based on the NLO properties of PYP films and that the optical switching efficiency can be manipulated by creating a steady state of PYP intermediates.

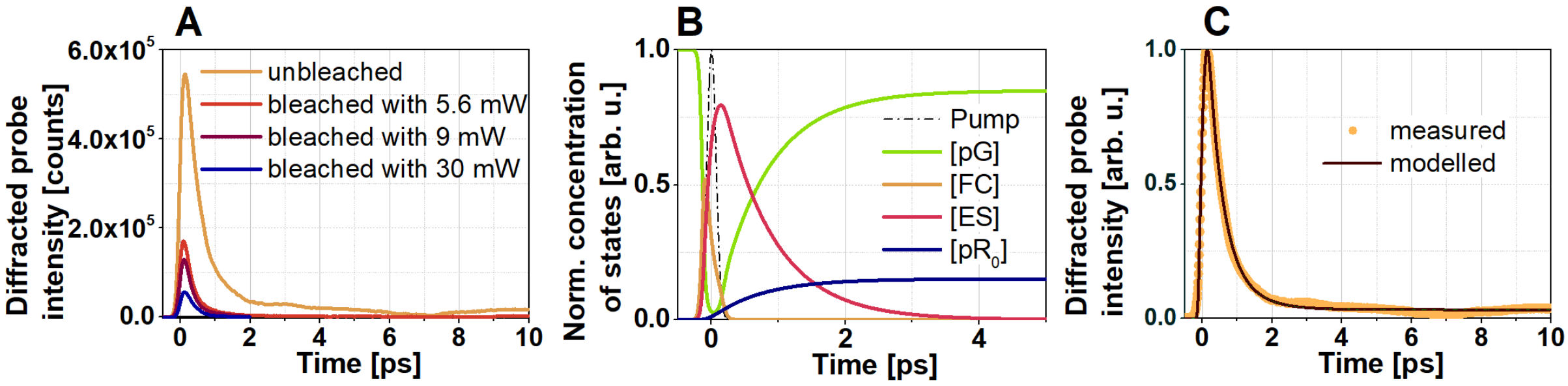

**FIG. 3.** Dynamics of glycerol doped thin PYP films. **(A)** Diffracted probe intensity as a function of time at a probe wavelength of 514 nm. The 120-fs pump pulses were centered at 450 nm. **(B)** Simulation of the normalized concentrations of the pG, FC, ES and $pR_0$ intermediates. The temporal shape of the pump pulse is shown by the dash-dot line. **(C)** Measured and calculated diffracted probe intensity as a function of time.

Based on the literature, only state pG and $pR_0$ contribute to the absorption change at 514 nm within the first 10-ps time frame.[28,34,52,67] However, using a model of the time-dependent diffracted probe intensity assuming only the presence of the two early intermediate states, pG and $pR_0$, the measured data could not be reproduced. Gradually extending the number of states in the model by adding a Franck-Condon and one or more excited states (similarly to that of the photocycle of PYP in liquid state[34,68]) and performing data regression based on the least squares method and Eq. (1), at least four states and four processes were needed for a satisfactory fit:

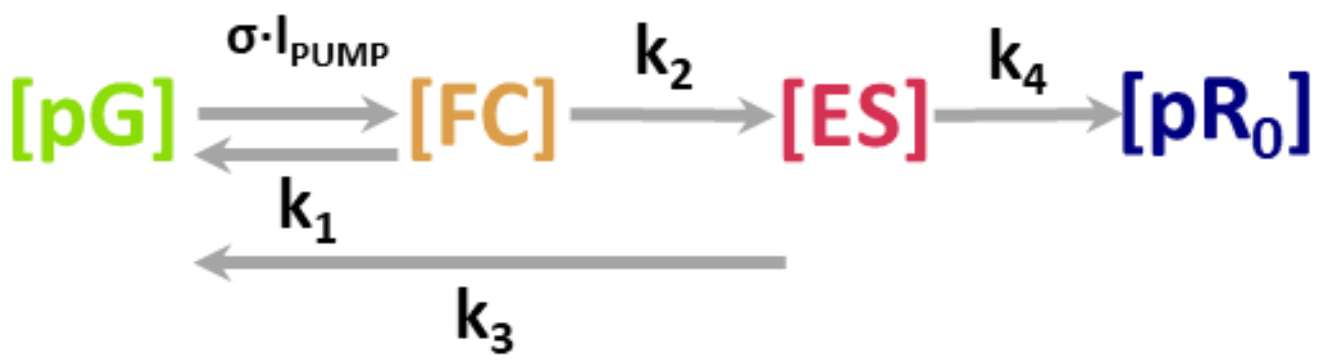

**Scheme 3** Kinetic scheme for rate equations describing the initial steps of the photocycle of photoactive yellow protein.

Here, pG, FC, and ES correspond to the concentration of the ground state, the Franck-Condon state, and the excited state, respectively. Satisfactory fit is obtained when mainly states ES and $pR_0$ contribute to the diffracted signal. $k_{1\text{-}4}$ are the rate constants, $I_{PUMP}$ is the applied pump intensity, and $\sigma = 1.682 \cdot 10^{-16}$ cm$^2$ is the absorption cross-section[68] of the ground state at 450 nm. Figure 3B shows the so-obtained time-dependent concentrations, while the measured and modelled diffracted probe signals are presented in Fig. 3C.

For the rate constants, we obtained $k_1$=1/42 fs$^{-1}$, $k_2$=1/96 fs$^{-1}$, $k_3$=1/876 fs$^{-1}$, and $k_4$=1/5364 fs$^{-1}$. Based on these results, we propose Scheme 3 for the early state of the photocycle model for thin, hydrated PYP films at an excitation wavelength of 450 nm.

The film dopant material had minor influence on the switching speed of PYP films. In contrast, shifting the excitation wavelength from the absorption peak of the pG state at 450 nm towards the red side, 480 nm, led to a drastic shortening of the switching time. With GL-doped films, the switched diffraction signal drops to 50% of its peak value at a delay 160 fs with 480-nm excitation, as opposed to 520 fs obtained with a 450-nm pump. In addition, instead of a delayed buildup of the diffraction signal relative to the temporal peak of the excitation pulse as was seen with 450-nm pumping and glycerol doping (i.e., 140 fs), we obtained a negligible lag of 20 fs with 480-nm excitation. Figure 4A and 4B show the corresponding measured diffraction intensity up to a delay of 10 and 2 ps, respectively, summarizing these observations. Modelling of the time-dependent diffraction signal for the GL-doped film at a pump wavelength of 480 nm led to the rate constants, $k_1$=1/20 $fs^{-1}$, $k_2$=1/170 $fs^{-1}$, $k_3$=1/1500 $fs^{-1}$, and $k_4$=1/4040 $fs^{-1}$ (cf. Fig. 4C). The rate values suggest that the much faster rise and decay time are due to the faster response of the Franck-Condon state at 480 nm than at 450 nm. We attribute the faster switching speed at 480 nm than at 450 nm to a higher stimulated emission cross section and concomitant emission, which forces the protein back to its ground state, as was also observed by Kuramochi et al[28].

The initial, few-100-fs and the final, 10-ps kinetics of the PAM-doped films followed closely that of the GL-doped films and showed somewhat slower decay components only on the intermediate few 100 fs – few ps time scale (cf. Fig. 4A and 4B, lines versus symbols). We tentatively attribute this difference between GL and PAM-doped films to a difference in the relative humidity and the protein structural heterogeneity.

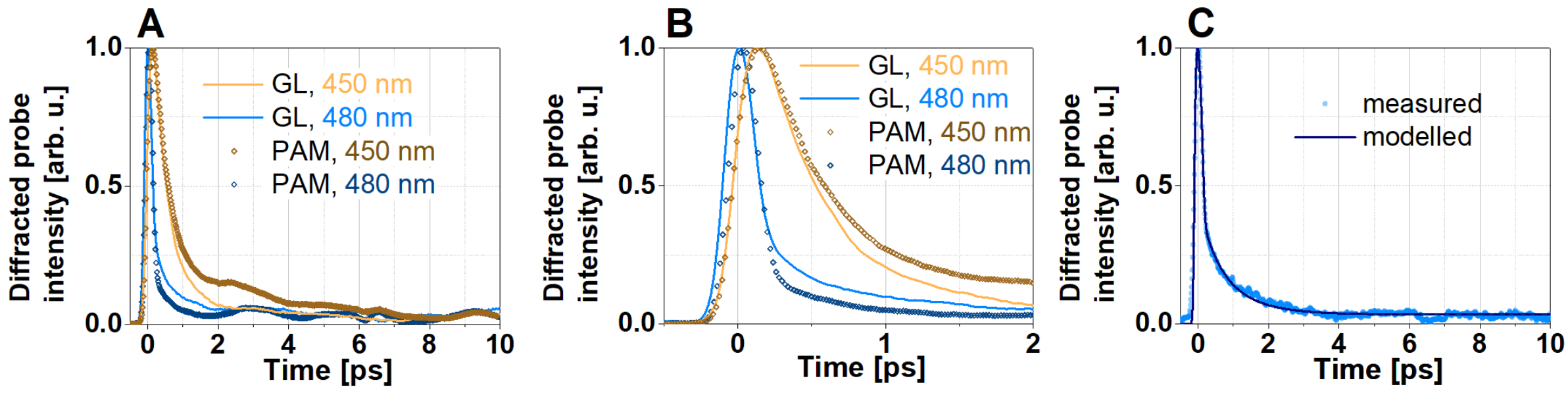

**FIG. 4.** Dynamics of PYP films. (A) Diffracted probe intensity as a function of time at a probe wavelength of 514 nm for GL-doped (solid lines) and PAM-doped (symbols) films. The excitation pulses were centered at 450 nm (yellow) or 480 nm (blue). (B) Zoomed-in version of (A) showing the temporal dependence up to 2 ps. (C) Measured and calculated diffracted 514-nm probe intensity as a function of time obtained for the GL-doped film with 480-nm, 160-fs excitation.

## Summary and outlook

Employing the folded BOXCARS geometry of third-order nonlinear optical spectroscopy, we demonstrated ultrafast all-optical switching with doped PYP films for the first time, with a rise and fall time down to that of the 160-fs pump laser pulse depending on the excitation wavelength. We used glycerol and acrylamide as doping materials and monitored the temporal evolution of the laser-induced concentration grating in the films up to a delay of 10 ps. The film dopant material had minor influence on the switching speed of PYP films. In contrast, the excitation wavelength had a drastic effect on the switching time possibly due to the varying contribution of stimulated emission from the Franck-Condon to the ground state. Using coupled rate equations, we modelled the experimental time-dependent diffracted probe intensity and found that an extension of the early-state photocycle model of PYP was needed to obtain a better fit between experiment and theory. In contrast, our experimental benchmark data on doped bacteriorhodopsin films could be fully explained using the known photocycle scheme for the known early intermediates. Our results show that an all-optical logic component based on PYP can be envisioned on the sub-ps timescale, enabling THz switching speed.

It is important to note that the ultrafast TG experiments, we report in this paper, go beyond our earlier results of all-optical on/off switching demonstrated by using bR films[25] and suggest the feasibility of a router-type all-optical packet switching by both chromoproteins (BR and PYP), as well, where data trains can be alternatively switched between different routes. While on/off switching can be used for encoding data, router-type switching is needed typically in Internet nodes, in order to deliver optically coded messages to the right address.[3,69] The latter requires a latching-type switch, where the on-rate should be extremely fast, and it should stay on during the transport of the whole data train ("packet").[3,69] In our experiments, the switch-on rate is following the duration of the excitation pulse (i.e., ca. 100 fs between 10% and 90% of maximal diffracted light intensity), while the switch stays in the "on" state until the "full" decay of the diffracted signal (for ca. 1 ps). We also showed that these switches can be temporarily inactivated by proper background CW illumination. It remains to be seen that the duration of the "on" state can be extended by using longer excitation pulses. Finally, the excited-state concentration grating employed in our scheme may also be used as an optically controlled diffractive element in future wavelength-selective switches (WSS).

For practical applications, the desirable high repetition rate remains to be demonstrated. Even though the repetition rate was limited by the duration of the photocycles of the chromoproteins in the present experiments, there are at least two straightforward solutions to the problem. One of them is utilizing the well-known photosensitivity of the intermediates in the photocycles of both BR and PYP,[39,66,70] namely, excitation of the intermediate states drives the proteins back to their ground-state conformations. In other words, the photocycles of both proteins are programmable by light. The other opportunity is a chemical modification of the chromophores. It has been demonstrated that both the retinal in bR and the p-coumaric acid in PYP can be removed by hydroxylamine treatment,[71–73] and the apoproteins can subsequently be reconstituted by non-isomerizable analogues of both chromophores.[36,74] In such chemically modified chromoproteins, only the Franck-Condon state and the quasi-stationary excited-state intermediate (I and ES, respectively) are allowed to be formed, because formation of the rest of the intermediates requires isomerization.[36,75] The I and ES excited states then spontaneously return to the ground states in the course of 10 ps (optimally suited for larger data packets).

Since the production of chromoproteins is cheap and they have high enough refractive index and absorption change, such proteins could be good candidates not only for light switching, but also for building holographic memories or image filtering.[76,77] In addition, BR and PYP have complementary advantages: BR works in a somewhat broader spectral range (between ca. 400 and 650 nm) as compared to PYP (between ca. 360 and 500 nm), while PYP, being a relatively small, water-soluble protein compared to BR that is available embedded in large membrane fragments, can be easily combined with nanostructured devices which makes it a suitable candidate as an active element in e.g., nanostructured IO devices.[27,78]

## Acknowledgements

The authors are indebted to Prof. Hinorari Kamikubo for kindly providing the plasmid for PYP expression, and to Dr. Tomás Zakar for the protein preparation. The work has received funding from the National Research, Development and Innovation Office, Hungary (NKFI-1 K-124922), the Eotvos Lorand Research Network (ELKH KÖ-36/2021), and the Deutsche Forschungsgemeinschaft (DFG, No. GSC 1013 SALSA). Z. H. acknowledges funding by a Julia Lermontova Fellowship from DFG (GSC 1013 SALSA). S. K. acknowledges funding by the German Academic Exchange Service (DAAD) and the Eotvos Hungarian State Scholarship of Tempus Public Foundation funded by the Hungarian Government.

## Author Declarations

The authors have no conflicts to disclose.